\documentclass[prd,preprint,double-spaced,superscriptaddress,showpacs]{revtex4-1}
\begin{document}


\title{Using Time Drift of Cosmological Redshifts to find the Mass-Energy Density of the Universe} 

\date{\today}

\author{M.E. Ara\'{u}jo}
\affiliation{ Departamento de F\'{\i}sica-Matem\'{a}tica, Instituto de F\'{\i}sica,\\
    Universidade Federal do Rio de Janeiro,\\
            21.945-970, Rio de Janeiro, R.J., Brazil}
\author{W.R. Stoeger}
\affiliation{ Vatican Observatory Research Group \\
      Steward Observatory, University of Arizona, \\
      Tucson, AZ 85721, USA}

\begin{abstract}

In this paper we show that the mass-energy density of the Universe can be fully determined in 
terms of the cosmological redshifts, their time drifts and angular-diameter distance (observer area distance). Besides  providing an indirect measurement of the mass-energy density of the Universe,  we show how one can use the time-drift of the cosmological redshifts as a replacement for the
mass-energy density element in the minimally required data set to construct an 
spherically symmetric Lema\^{\i}tre-Tolman-Bondi (LTB) model for the Universe in observational coordinates.

\end{abstract}


\maketitle

\section{Introduction}

The present astrophysical data sets show that the Universe is almost isotropic around us on the largest scales. Using this observed isotropy  the Copernican principle leads to a Universe modeled by the Friedmann-Lema\^{\i}tre-Robertson-Walker (FLRW) space-time geometry. Relaxing the assumption that the Copernican principle holds, the Universe is then modeled by the spherically-symmetric LTB space-time geometry.

The primary aim of OC program is to strengthen the connections between astronomical observations and cosmological theory. We do this by allowing observational data to determine the geometry of space-time as much as possible, {\it without} relying on {\it a priori}
assumptions more than is necessary or justified. Basically, we want to find out not only how far our observable Universe is from being isotropic and spatially homogeneous (that, is describable by an FLRW cosmological model) on various length scales, but also to give a dynamic account of those deviations (Stoeger {\it et al} \cite{OC III}).

Recently, Ara\'ujo and Stoeger \cite{AS2009b}, showed how to construct all spherically symmetric, inhomogeneous cosmological (LTB) models in observational coordinates from cosmological data on our past light cone, consisting of galaxy redshifts, luminosity distances (observer area distances) and mass-energy densities (or, galaxy number counts) as functions of redshift and allowing for a nonzero cosmological constant (vacuum energy). In doing so we provided a new rigorous demonstration of how such data fully determines the time evolution of all the metric components, and a Taylor series algorithm for determining those solutions. This enables us to move the solution we obtain from data on the light cone off it in a well-defined and straightforward way. It is essential for this to have data giving the maximum of  the observer area (angular-diameter) distance, $C_0(w_0, z_{\max})$, and the redshift $z_{max}$ at which that occurs. This enables the determination of the vacuum-energy density $\mu_{\Lambda}$, which would otherwise remain undetermined.

One of the key issues concerning observations to acquire real data is the determination of the mass-energy density. For that -- if we rely on galaxy number counts -- we need an accurate determination of the average mass per galaxy as a function of redshift -- which to be reliable, requires an appropriate model of galaxy evolution which, in turn, requires a precise determination of the cold dark matter content in galaxies. 

Here we show that the mass-energy density of the Universe can be fully determined in 
terms of the cosmological redshift, the time drift of the cosmological redshift, hereafter 
redshift-drift, and the observer area distance. Besides its importance in its own right, 
providing an indirect measurement of the mass-energy density
of the Universe, thus  bypassing all the above mentioned problems related to a direct measurement, we can use the redshift-drift as a replacement for
mass-energy density element in the minimally required data set needed to construct an LTB model for the Universe in observational coordinates. Moreover, our result is strengthened by noting that  the determination of the mass-energy density is completely independent of the freedom we have in the observational coordinate formalism to specify the time coordinate $w$ along our world line $\cal C$ (see below). 

This relationship of the redshift-drift $\dot z$ to the mass energy-density was first obtained by Sandage \cite{sand} and McVittie \cite{mcv}, and expressed as

\begin{equation}
\dot z = (1+z) H_0 + H(z) \label{rdsm}
\end{equation}
where, $H_0$ is the Hubble parameter now and $H(z)$ is the Hubble parameter at redshift $z$. $H(z)$ is a function of the mass-energy density. Recently, Uzan, Clarkson and Ellis \cite{uce}, working in the $3+1$ framework, also showed that observation of redshift-drift would allow us to test the Copernican principle. The LTB spacetime
in the $3+1$ formalism depends on three arbitrary functions of the space-like radial coordinate 
$r$, one of which can be fixed by a gauge freedom in $r$. They use this freedom to fix $M(r)$,  a
quantity related to number counts (see below), and proceed to determine the other two using
the redshift-drift and observer area distance. The results proven here clearly show that in the OC formalism we do not need to fix $M$ in that way. Instead, we explicitly determine the mass 
energy density (number counts) as a function of the observer area distance and the redshift-drift.
Uzan, Clarkson and Ellis \cite{uce}  also claim that writing the LTB metric in observational coordinates requires the solution of the null geodesic equation, which is in general only possible numerically. Given the recent results of Ara\'ujo and Stoeger \cite{AS2009b} mentioned above and those shown in his paper, we find that this claim is an overstatement. 

Although we do not yet have the capability to measure redshift drift, the
prospects for doing so within the next 20 years are good. This should be able
to be done using the European Extremely Large Telescope (ELT) and its CODEX
ultra stable spectrograph (see \cite{Loeb} and \cite{Pasq}). Both of these are presently under development. This makes the results given here observationally significant.

In the next section we define observational coordinates, write the general spherically symmetric metric using them and present the central conditions for the metric variables. Section \ref{sec:obspar} presents the basic observational parameters we shall be using and
several key relationships among the metric variables. Section \ref{sec:fieqs} presents the full set of field equations for the spherically symmetric case, with dust and with $\Lambda \neq 0$. In section \ref{sec:tddata} we present our main result showing how the mass-energy density function can be expressed in terms of the redshift, the redshift-drift and the observer area distance. 
In light of this new set of data functions, Section \ref{sec:nonflat} reviews our general integration scheme for all inhomogeneous spherically symmetric LTB models. In section \ref{sec:concl} we briefly discuss our conclusions.

\section{ The Spherically Symmetric Metric in Observational Coordinates}

We are using observational coordinates (which were first suggested by
Temple \cite{Temple}). As described by Ellis {\it el al}  \cite{Ellis et al}  the observational
coordinates $x^i=\{w,y,\theta ,\phi \}$ are centered on the observer's
world line $ {\cal C}$ and defined in the following way: \\

\noindent
(i) $w$ is constant on each past light cone along $ {\cal C}$, with $u^a
\partial _a w > 0$ along $ {\cal C}$, where $u^a$ is the 4-velocity of matter
($u^au_a=-1$). In other words, each $w = constant$ specifies a past
light cone along $ {\cal C}$. Our past light cone is designated as $w =
w_0$. \\

\noindent
(ii) $y$ is the null radial coordinate. It measures distance down the null
geodesics -- with affine parameter $\nu$ -- generating each past light cone
centered on $ {\cal C}$. $y = 0$ on $ {\cal C}$ and $dy/d\nu > 0$ on each null cone -- so
that $y$ increases as one moves down a past light cone away from $ {\cal C}$. \\

\noindent
(iii) $\theta$ and $\phi$ are the latitude and longitude of 
observation, respectively -- spherical coordinates based on a
parallelly propagated orthonormal tetrad along $ {\cal C}$, and defined away
from $ {\cal C}$ by $k^a \partial _a \theta = k^a \partial _a \phi = 0$, where
$k^a$ is the past-directed wave vector of photons ($k^ak_a=0$). \\

\noindent
There are certain freedoms in the specification of these observational
coordinates. In $w$ there is the remaining freedom to specify $w$ along our
world line $ {\cal C}$. Once specified there it is fixed for all other world lines.
There is considerable freedom in the choice of $y$ -- there are a large
variety of possible choices for this coordinate -- the affine parameter, the red-shift,
the luminosity distance. We normally choose $y$ to be comoving
with the fluid, that is $u^a\partial _ay=0$. Once we have made this choice,
there is still a little bit of freedom left in $y$, which we shall use below.
The remaining freedom in the $\theta $ and $\phi $ coordinates is a rigid
rotation at $one$ point on $ {\cal C}$. 

In observational coordinates the Spherically Symmetric metric takes
the general form:
\begin{equation}
ds^2=-A(w,y)^2dw^2+2A(w,y)B(w,y)dwdy+C(w,y)^2d\Omega ^2,  \label{oc}
\end{equation}
where we assume that $y$ is comoving with the fluid, so that the
fluid 4-velocity is $u^a=A^{-1}\delta _w^a$.

The remaining coordinate freedoms which preserve the observational form of
the metric are scalings of $w$ and of $y$:
\begin{equation}
w\rightarrow \tilde{w}=\tilde{w}(w)~,~~y\rightarrow\tilde{y}= \tilde{y}%
(y)~~~~\left({\frac{d\tilde{w}}{dw}}\neq 0 \neq {\frac{d\tilde{y} }{dy}}%
\right).  \label{wy}
\end{equation}

The first, as we mentioned above, corresponds to a freedom to choose $w$ as
any time parameter we wish along $ {\cal C}$, our world line at $y=0$. This is
usually effected by choosing $A(w,0)$. For instance, if the coordinate $w$ is normalized 
by the condition that it measures proper time along ${\cal C}$, then $A(w,0)=1$. 
The second corresponds to the freedom to choose $y$ as any null distance parameter 
on an initial light cone -- typically our light cone at $w=w_0$. 
Then that choice is effectively dragged onto other light cones by the fluid flow. 
Thus we here choose $y$ to be co-moving with the fluid
4-velocity, as we have already indicated. We shall use this freedom to
choose $y$ by setting:
\begin{equation}
A(w_0, y) = B(w_0, y).  \label{ab}
\end{equation}
We should carefully note here that setting $A(w, y) = B(w, y)$ off our
past light cone $w = w_0$ is too restrictive.

In general, these freedoms in $w$ and $y$ imply the metric scalings:
\begin{equation}
A\rightarrow\tilde{A}={\frac{dw}{d\tilde{w}}}A~,~~ B\rightarrow\tilde{B}={%
\frac{dy}{d\tilde{y}}}B.  \label{4scale}
\end{equation}

It is important to specify the central conditions for the metric variables $%
A(w, y)$, $B(w, y)$ and $C(w, y)$ in Eq. (\ref{oc}) -- that is, their proper
behavior as they approach $y = 0$. These are:
\begin{eqnarray}
{\rm as}\;\;y\rightarrow 0:\;\;\; &&A(w,y)\rightarrow A(w,0)\neq 0, 
\nonumber \\
&&B(w,y)\rightarrow B(w,0)\neq 0,  \nonumber \\
&&C(w,y)\rightarrow B(w,0)y = 0,  \label{cent} \\
&&C_y(w,y)\rightarrow B(w,0).  \nonumber
\end{eqnarray}
These important conditions insure that ${\cal C}$, our world line, is
regular -- so that all functions on it are bounded, and that the
spheres ($w$, $y =$ constant) go smoothly to ${\cal C}$ as $y \rightarrow 0$.
They also insure that the null surfaces $w =$ constant  are past light cones
of observers on ${\cal C}$ (See reference  \cite{Ellis et al} , especially section 3.2, p.
326, and Appendix A for details).

\section{\label{sec:obspar}The Basic Observational Quantities}

The basic observable quantities on $ {\cal C}$ are the following: \\

(i) Redshift. The redshift $z$ at time $w_0$ on $ {\cal C}$ for a co-moving source a
null radial distance $y$ down our past light cone $C^{-}(p_0)$, which is centered at our observational position $p_0$, is given by
\begin{equation}
1+z={\frac{A(w_0,0)}{A(w_0,y)}}.  \label{z}
\end{equation}
This is just the observed redshift, which is directly determined by source
spectra, once they are corrected for the Doppler shift due to local motions.
It is consistent with and complements  the first of the central conditions in
Eq. (\ref{cent}). 

We can generalize Eq. (\ref{z}) for the redshift $z$ at an arbitrary time $w$ on $ {\cal C}$ for a comoving source a null radial distance $y$ down $C^{-}(p)$  by
\begin{equation}
1+z(w,y) ={\frac{A(w,0)}{A(w,y)}}.  \label{zgen}
\end{equation}

 (ii) Redshift-drift (time drift of the cosmological redshift). There are the two fundamental four-vectors
in the problem, the fluid four-velocity $u^a$ and the null vector $k^a$, which points down the
generators of past light cones. These are given in terms of the metric
variables as
\begin{equation}
u^a = A^{-1}\delta^a{}_w ~,~~ k^a = (AB)^{-1}\delta^a{}_y.  \label{uk}
\end{equation}
From the normalization condition for the fluid four-velocity, we
can immediately see that it can be given (in covariant vector form) as the
gradient of the proper time $\tau$ along the matter world lines: $u_a=-\tau,_a$.
It is also given by (\ref{oc}) and (\ref{uk}) as
\begin{equation}
u_a=g_{ab}u^b=-Aw_{,a}+By_{,a}.
\end{equation}
Comparing these two forms implies
\begin{equation}
d\tau=Adw-Bdy~~\Leftrightarrow~~A=\tau_w~,~~ B=-\tau_y,  \label{tw}
\end{equation}
which shows that the surfaces of simultaneity for the observer are given in
observational coordinates by $A dw = B dy$. 
The integrability condition of Eq. (\ref{tw}) is simply then

\begin{equation}
A^{\prime}+\dot{B}=0.  \label{coneq}
\end{equation}
where a ``dot'' indicates $\partial/\partial w$ and a ``prime'' indicates $\partial/\partial y$. 
This turns out precisely to be the momentum conservation equation, which is
a key equation in the system and essential to finding a solution. 

It follows from Eq. (\ref{tw}) that

\begin{equation}
\frac{d\tau}{dz}=A\frac{dw}{dz} - B\frac {dy}{dz}=\frac {A}{\dot z} - B\frac {dy}{dz}. \label{dtdz}
\end{equation}
where, $\dot z \equiv {\partial z}/{\partial w}$ is the redshift-drift. We also note, in passing, that
we can write ${\partial w}/{\partial z} = {dw}/{dz}$ and ${\partial z}/{\partial y} = {dz}/{dy}$ because
 $w$ and $y$ are independent coordinates.

It is clear that on the radial null geodesics $\tau = \tau (w,y)$ and $z=z(w,y)$. Now, since on our 
past light cone  $w=w_0$ -- in OC coordinates $w$ labels past light  cones -- we must have that

\begin{equation}
d\tau  = k(y) dz \label{tauz}
\end{equation}
where, $k(y) \ne 0$.

Hence, it  follows from Eq. (\ref{tauz})  that Jacobian of the transformation $\{\tau, z\}\mapsto \{w,y\}$ must vanish on our past light cone. Since,

\begin{equation}
\tilde J= \frac {\partial (\tau,z)} {\partial(w,y)} = \left|
\begin{array}{c c } \frac {\partial \tau} {\partial w} &  \frac {\partial z} {\partial w} \\  \frac {\partial \tau} {\partial y}  &  \frac {\partial z} {\partial y} \end{array} \right| = A(w,y)\frac {\partial z } {\partial y} +  B(w,y) {\dot z}  \label{jacob}
\end{equation}
we  have that 

\begin{equation}
\tilde J(w_0,y)= 0 \Rightarrow \frac{B(w_0,y)}{A(w_0,y)}= -\frac{1}{\dot z}\frac{\partial z}{\partial y}, 
\label{jacob0}
\end{equation}





Since ${\dot z(z)}$ is given from data, solving Eq. (\ref{jacob0}) gives $z=z(w_0,y)$. That is the same information that would be obtained from the null Raychaudhuri [Eq. (\ref{nr}) below]. It is quite clear 
that, whereas to solve the later on our past light cone one needs the 
mass-energy density $\mu(w_0,z)$, what we need to solve Eq. (\ref{jacob0}) is a different piece
of information -- the redshift-drift  $\dot z(w_0,z)$. Moreover, we shall demonstrate later in this paper that the mass-energy density $\mu_{m_0}(z)$ can be completely determined in terms of the redshift, the redshift-drift and the observer area distance $C(w_0,z)$ data on our past light cone.

(iii) Observer Area Distance. The observer area distance, often written as $
r_0$, measured at time $w_0$ on $ {\cal C}$ for a source at a null radial distance $y$
is simply given by
\begin{equation}
r_0=C(w_0,y),
\end{equation}
provided the central condition (\ref{cent}), determining the relation
between $C(w,y)$ and $B(w,y)$ for small values of $y$, holds. This quantity
is also measurable as the luminosity distance $d_L$ because of the reciprocity
theorem of Etherington \cite{Etherington33} (see also Ellis \cite{Ellis 1971}),\\ 

\begin{equation}
d_L = (1+z)^2 C(w_0, y).   \label{recth} \\
\end{equation}

(iv) The Maximum of Observer Area Distance. Generally speaking, $C(w_0, y)$
reaches a maximum $C_{max}$ for a relatively small redshift $z_{max}$ (Hellaby
 \cite{Hellaby}; see also Ellis and Tivon  \cite{ET} and Ara\'{u}jo and Stoeger  \cite{ASII}). At
$C_{max}$, of course, we have 

\begin{equation}
\frac{d C(w_0, z)}{d z} = \frac{d C(w_0, y)}{d y} = 0,\\
\end{equation}
further conditioned by
\begin{equation}
\frac{d^2 C(w_0, z)}{d z^2} < 0.\\
\end{equation}
Furthermore,  with the solution of Eq. (\ref{jacob0}), 
the redshift-drift  will give us $y = y(z)$, from which we shall be
able to find $y_{max} = y_{max}(z_{max})$.  These $C_{max}$ and $z_{max}$
data provide additional independent information about the cosmology. Without
$C_{max}$ and $z_{max}$ we cannot constrain the value of $\Lambda$. 
 
There are a number of other important quantities which we catalogue here for
completeness and for later reference. 

Galaxy Number Counts. The number of galaxies counted by a central
observer out to a null radial distance $y$ is given by
\begin{equation}
N(y)=4\pi\int_0^y \mu(w_0,\tilde{y})m^{-1}B(w_0,\tilde{y})C(w_0,\tilde{y})^2
d\tilde{y},  \label{N}
\end{equation}
where $\mu$ is the mass-energy density and $m$ is the average galaxy mass.
Then the total energy density can be written as
\begin{equation}
\mu(w_0,y) = m\;n(w_0,y) = M_0(z)\;{\frac{dz}{dy}}\;{\frac{1}{B(w_0,y)}},
\label{mudef}
\end{equation}
where $n(w_0, y)$ is the number density of sources at $(w_0, y)$, and where
\begin{equation}
M_0 \equiv {\frac{m}{J}}\;{\frac{1}{d\Omega}}\;{\frac{1}{r_0^2}}\;{\frac{dN}{dz}}. \label{m0def}
\end{equation}
Here $d \Omega$ is the solid angle over which sources are counted, and
$J$ is the completeness of the galaxy counts, that is, the fraction of
sources in the volume that are counted is $J$. The effects of dark
matter in biasing the galactic distribution may be incorporated via $m$ and/or
$J$ . In particular, strong biasing is needed if the number counts have
a fractal behavior on local scales (Humphreys {\it et al} \cite{hmm}). In order
to effectively use number counts to constrain our cosmology, we shall also
need an adequate model of galaxy evolution. We shall not discuss this
important issue in this paper. But, fundamentally, it would give us
an expression for $m = m(z)$ in Eqs. (\ref{mudef}) and (\ref{m0def}) above.

The Hubble parameter. The rate of expansion of the dust fluid is $3 H = \nabla_a u^a$, so that,
from the metric (1) we have:
\begin{equation}
H={\frac{1}{3A}}\left({\frac{\dot{B}}{B}}+2{\frac{\dot{C}}{C}}\right),
\label{h}
\end{equation}
For the central observer $H$
is precisely the Hubble expansion rate. In the homogeneous (FLRW) case, $H$ is
constant at each instant of time t. But in the general inhomogeneous case, $%
H $ varies with radial distance from $y = 0$ on $t = t_0$. From our central
conditions above (3), we find that the central behavior of $H$ is given by
\begin{equation}
{\rm as}\;\;y\rightarrow 0:\;\;\;H(w,y)\rightarrow {\frac{1}{A(w,0)}}{\frac{%
\dot{B}(w,0)}{B(w,0)}}=H(w,0).  \label{hcent}
\end{equation}
At any given instant $w = w_0$ along $y = 0$, this expression is just the
Hubble constant $H_0 \equiv H(w_0, 0) = A_0^{-1} B_0^{-1}(\dot{B})_0$ as
measured by the central observer. In the above we have also written $A_0 \equiv A(w_0, 0)$
and  $B_0 \equiv B(w_0, 0)$.

\section{\label{sec:fieqs}The Spherically Symmetric Field Equations in 
Observational Coordinates}

Using the fluid-ray tetrad formulation of the Einstein's equations 
developed by Maartens \cite{m} and Stoeger {\it et al} \cite{fluid ray}, 
one obtains the Spherically Symmetric field equations in observational 
coordinates with $\Lambda \neq 0$ (see Stoeger {\it et al} \cite{OC III} for a
detailed derivation). Besides the momentum conservation Eq. (\ref{coneq}), they are
as follows:

A set of two very simple fluid-ray tetrad time-derivative equations:

\begin{eqnarray}
\dot{\mu}_m &=& -2{\mu_m} \left(\frac{\dot B}{2B} + \frac{\dot{C}}{C} \right), \label{mueqr} \\
\dot{\omega} &=& -3 \frac{\dot C}{C} \biggl(\omega + \frac{\mu_{\Lambda}}{6} \biggr), \label{omegaeqr}
\end{eqnarray}
where $\mu_m$ again is the relativistic mass-energy density of the dust, 
including dark matter, and 

\begin{equation}
\omega(w,y)\equiv -{\frac{1}{2C^2}} + {\frac{
\dot{C}}{{AC}}}{\frac{C^{\prime}}{{BC}}} + {\frac{1}{2}} \biggl({\frac{
C^{\prime}} {BC}}\biggr)^2, \label{omegadef}
\end{equation}
is a quantity closely related to $\mu_{m_0}(y)\equiv \mu_m(w_0,y)$ (see Eq.\ref{omegdef}) below).

Equations (\ref{mueqr}) and (\ref{omegaeqr}) can be quickly integrated to give:
\begin{widetext}
\begin{equation}
\mu_m(w,y)=\mu_{m_0}(y)\;{\frac {B(w_0,y)} {B(w,y)}}\; \frac {C^{2}(w_0,y)} {C^{2}(w,y)}; 
\end{equation}

\begin{equation}
\omega(w,y)=\biggl(\omega_0(y) + {\frac {\mu_{\Lambda}} {6} }\biggr) {\frac{C^3(w_0, y)}{C^3(w,y)}}
- {\frac {\mu_{\Lambda}} {6}} = {-{\frac{1}{{2C^2}}}+{\frac{
\dot{C}}{{AC}}}{\frac{C^{\prime}}{{BC}}}+{\frac{1}{2}}\biggl({\frac{
C^{\prime}}{{BC}}}\biggr)^2}, \label{omega} 
\end{equation}
\end{widetext}
where $\omega_0(y)\equiv \omega(w_0,y)$ and the last equality in (\ref{omega}) follows from the definition of $\omega$ given above. In deriving and solving these equations, and those below, we have used the
typical $\Lambda$ equation of state, $p_{\Lambda} = - \mu_{\Lambda},$
where $p_{\Lambda}$ and $\mu_{\Lambda} \equiv \frac{\Lambda}{8 \pi G}$ are the pressure and the 
energy density due to the cosmological constant. Both $\omega_0$ and
$\mu_0$ are specified by data on our past light cone, as we shall show.
$\mu_{\Lambda}$ will eventually be determined from the measurement of $C_{max}$
and $z_{max}$. 

The fluid-ray tetrad radial equations are:
\begin{widetext}
\begin{eqnarray}
&{\frac{C^{\prime\prime}}{C}} = {\frac{C^{\prime}}{C}}{\biggl({\frac{
A^{\prime}}{A}} +{\frac{B^{\prime}}{B}}\biggr)} - {\frac{1}{2}}B^2\mu_m;
\label{nr} \\
&\biggl[ \bigl(\omega_0(y) + {\frac {\mu_{\Lambda}} {6} }\bigr)C^3(w_0, y)\biggr]^{\prime} = -{\frac{1}{2}}\mu_{m_0}\;
{B(w_0,y)}\;{C^{2}(w_0,y)}\;{\biggl({\frac{\dot {C}}{A}} +
 {\frac{C^{\prime}}{B}}\biggr)};  \label{omegap} \\
&{\frac{{\dot {C}}^{\prime}}{C}} = {\frac{{\dot B}}{B}}{\frac{C^{\prime}}{C}
} - \biggl(\omega + {\frac {\mu_{\Lambda}} {2}}\biggr)\; A\;B.  \label{prdot}
\end{eqnarray}
\end{widetext}
The remaining ``independent'' time-derivative equations given by the
fluid-ray tetrad formulation are:

\begin{eqnarray}
&&{\frac{{\ddot{C}}}C}={\frac{{\dot{C}}}C}{\frac{{\dot{A}}}A}+ \biggl(\omega + 
{\frac {\mu_{\Lambda}} {2}}\biggr) \;A^2;
\label{Cdd} \\
&&{\frac{{\ddot{B}}}B}={\frac{{\dot{B}}}B}{\frac{{\dot{A}}}A}-2\omega \;A^2-{%
\frac 12}\mu_{m} \;A^2 .  \label{bdd}
\end{eqnarray}
From Eq. (\ref{omegap}) we see that there is a naturally defined
``potential'' (see Stoeger {\it et al} \cite{OC III}) depending only on the radial
null coordinate $y$ -- since the left-hand-side depends only on $y$, the
right-hand-side can only depend on $y$:

\begin{equation}
F(y)\equiv {\frac{\dot{C}}A}+{\frac{C^{\prime }}B},  \label{f1}
\end{equation} 
Thus, from Eq.  (\ref{omegap}) itself
\begin{widetext}
\begin{equation}
\omega_0(y)= - {\frac {\mu_{\Lambda}} {6} }- {\frac{1}{2 C^3(w_0, y)}}\;\int{\mu_{m_0}(y)\;
{B(w_0,y)}\;{C^{2}(w_0,y)}\;F(y)\;dy}.  \label{omegdef}
\end{equation}
\end{widetext}

We now quote a simple but very important observational relationship 
originally obtained by Hellaby  \cite{Hellaby} in the 3+1 framework [for a detailed
derivation of this result in observational coordinates see Ara\'ujo and Stoeger \cite{AS2009b}] :

\begin{equation} 
6M_{max} + \mu_{\Lambda} C^3_{max} - 3 C_{max} = 0 .\label{hellabyeq}
\end{equation}
 where,  $M(y)$ is a quantity given by


\begin{equation}
M(y) = \frac {1}{8 \pi} \int_0^y mN^{\prime}(\tilde y) F(\tilde y) d\tilde y  =  \frac {1}{8 \pi} \int_0^y \bar M^{\prime}(\tilde y) F(\tilde y) d\tilde y . \label{MN} 
 \end{equation}
and $M_{max}\equiv M(y_{max})$. We note that $ \bar M(y) = m N(y)$ is  the total mass summed over the whole sky by a central observer out to a null radial distance $y$.

Eq. (\ref{hellabyeq}) has to be considered a fundamental relation in Observational Cosmology, since it enables, from $C(w_0, z_{max})$ and $z_{max}$ measurements,  the determination of the unknown constant $\mu_{\Lambda}$.
  
\section{\label{sec:tddata} Mass density as a function of redshift, redshift drift and area distance}


Differentiating  Eq. (\ref{zgen})  with respect to $w$ gives

\begin{equation}
\dot z(w,y) = \frac{A(w,0)}{A(w,y)} \Biggl\{\frac{\dot A(w,0)}{A(w,0)} -  \frac{\dot A(w,y)}{A(w,y)} \Biggr\} = (1+z) \Biggl\{\frac{\dot A(w,0)}{A(w,0)} -  \frac{\dot A(w,y)}{A(w,y)} \Biggr\}.  \label{dzgen}
\end{equation}


Equation  (\ref{bdd}) can be rewritten as

\begin{equation}
\frac {A}{B} \frac {\partial}{\partial w} \Biggl(\frac{\dot{B}} {A}\Biggr) = -2\omega \;A^2-{
\frac 12}\mu_{m} \;A^2 . \label{bddm}
\end{equation}

Substitution of equation  (\ref{coneq}) into  (\ref{bddm}) yields

\begin{equation}
\frac {\partial}{\partial w}\Biggl( \frac{A^{\prime}} {A}\Biggl) =  \frac {\partial}{\partial w} \Biggl[\frac{\partial} {\partial y} (\ln A)\Biggr] = \frac {\partial}{\partial y} \Biggl[\frac{\partial} {\partial w} (\ln A)\Biggr] = \frac {\partial}{\partial y}\Biggl( \frac{\dot{A}} {A}\Biggr) = 2\omega \;AB + {\frac 12}\mu_{m} \;AB, \label{Adot}
\end{equation}
Hence, its general solution is

\begin{equation}
\dot{A}(w,y) = A(w,y)\Biggl\{ \int_0^y\Biggl[2\omega(w,\tilde y) +{
\frac 12}\mu_m(w,\tilde y)\Biggl] \;A(w,y)B(w,y) \;d{\tilde y} + l(w) \Biggr\}. \label{Adgen}
\end{equation}
where $l(w)$ is a function of $w$ given by

\begin{equation}
l(w)=\frac{\dot A(w,0)}{A(w,0)} \label{lw}
\end{equation}
It is clear from  Eq. (\ref{lw}) that $l(w)$ depends on the choice of the time coordinate $w$ on our world line.

Substituting Eqs. (\ref{Adgen}) and (\ref{lw}) into Eq. (\ref{dzgen})  gives

\begin{equation}
 \int_0^y\Biggl[2\omega (w,\tilde y) +{
\frac 12}\mu_{m}(w, \tilde y)\Biggl] \; A(w, \tilde y)B(w, \tilde y) \;d{\tilde y}=-\frac {\dot z(w,y)}{1+z(w,y)},  
\label{main0}
\end{equation}
It is important to observe at this point that  Eq.  (\ref{main0}) is independent of the function $l(w)$, thus showing that this result is manifestly independent of any particular choice of the time coordinate $w$. 

Differentiating Eq. (\ref{main0}) with respect to $y$ gives
\begin{equation}
\Biggl[2\omega (w,y) +{\frac 12}\mu_{m}(w,y)\Biggl]\;A(w, y)B(w, y)
= - \frac {\partial}{\partial y} \Biggl[ \frac {\dot z(w,y)}{1+z(w,y)}\Biggl].\label{main4}
\end{equation}
On our past light cone, $A(w_0, y) = B(w_0, y)$, and we find from Eq. (\ref{main4}) that

\begin{equation}
\mu_{m_0}(y) =  - \frac {2} {A^2_0(y)} \frac {\partial}{\partial y} \Biggl[ \frac {\dot z_0(y)}{1+z_0(y)}
\Biggl] -4\omega_0(y), \label{main5}
\end{equation}
where, we have denoted $A_0(y) \equiv A(w_0, y)$, and similarly used the same notation 
for all the other quantities.

Finally, using Eq. (\ref{jacob0}) and its solution, we can write  Eq. (\ref{main5}) in terms of the redshift $z$ as

\begin{equation}
\mu_{m_0}(z) =   \frac {2\dot z_0(z)} {A^2_0(z)} \frac {\partial}{\partial z} \Biggl[ \frac {\dot z_0(z)}{1+z_0(z)}
\Biggl] -4\omega_0(z). \label{main6}
\end{equation}

Eq.  (\ref{main6}) shows that the mass-energy density $\mu_{m_0}(z)$ can be completely determined in terms of the redshift $z$, the redshift-drift $\dot z(z)$ and observer area 
distance $C(w_0,z)$ data on our past light cone. We observe that the $\mu_{m_0}(z)$
dependency on the observer area distance $C(w_0,z)$, leads to its dependency on 
$\mu_{\Lambda}$, that must be determined by data giving the maximum of  the observer area distance, $C_0(w_0, z_{\max})$, and the redshift $z_{max}$ at which that occurs. In the next section we show how this is done.







Now, from Eqs.  (\ref{ab}),  (\ref{jacob0}) and (\ref{omegadef}) we find that

\begin{equation}
\omega_0(z) = -\frac{1}{2C_0^2(z)} \Biggl\{1+
\frac{ \dot z} {A^2_0(z)} \frac{dC_0(z)}{dz} \Biggl[2 \dot C_0(z) -  \dot z \frac{dC_0(z)}{dz}\Biggr]\Biggr\}
 \label{omega0}
\end{equation}
Clearly, $dC_0(z)/dz $ can be determined from the $r_0(z) \equiv C(w_0,z)$
data, through fitting. We determine $\dot{C}_0(z)$ by solving Eq. (\ref{prdot}) for  $\dot{C}_0(y)$ on $w = w_0$ [see Ara\'ujo and Stoeger \cite{AS2009b} for details] and then use Eq. (\ref{jacob0}) and its solution to write the result in terms of $z$. Taking into account the appropriate boundary condition (central condition),  $\dot{C}_0(z)$ is given by:

\begin{eqnarray}
\dot C_0(z) = - \frac {1}{2C_0(z)} & & \int_0^z \frac{1}{\dot z_0(\tilde z)}\Biggl \{A_0^2(\tilde z) 
- \dot z_0^2(\tilde z)\frac{dC_0(\tilde z)}{d \tilde z}
\Biggl[ \frac {2 C_0(\tilde z)}{ A_0(\tilde z)}\frac{dA_0(\tilde z)}{d \tilde z}+\frac{dC_0(\tilde z)}{d\tilde z}\Biggr] \nonumber \\
&  & \hfill{\quad}
 -A_0^2(\tilde z)C_0^2(\tilde z)\mu_{\Lambda} \Biggr\} d\tilde z. \label{coz}
\end{eqnarray}

Since, $\dot z(w_0,z)$ is very small, it follows very clearly from Eqs. (\ref{main6}), (\ref{omega0}) 
and (\ref{coz}) that a useful approximation to $\mu_{m_0}(z)$ is

\begin{equation}
\mu_{m_0}(z) \cong -4\omega_{0}(z).
 \label{main7}
\end{equation}

\section{\label{sec:nonflat}The General Solution - Time evolution off our Light Cone}

In this section we outline the general integration procedure that is applicable 
to all inhomogeneous spherically symmetric universe models  - that is the only constraint.  
[See Ara\'ujo and Stoeger \cite{AS2009b} for a detailed description of this procedure.]
We do not know whether the Universe is homogeneous  or not. But the data gives us 
redshifts $z$, observer area distances (angular-diameter distances) $r_0(z)$, 
time drift of cosmological redshifts $\dot z_0(z)$, and the angular-distance maximum
$C_{max}(w_0, z)$ at $z_{max}$. It is important to specify the latter, because,
as we have already emphasized, without them, we do not have enough information
to determine all the parameters of the space-time in the $\Lambda \neq 0$ case.
For instance, although we can determine $C(w_0, z)$ with good precision
(by obtaining luminosity distances $d_L$ and employing the reciprocity theorem,
equation  (\ref{recth})) out to relatively high redshifts, at present we do not yet have
reliable data deep enough to determine $C_{max}$ and $z_{max}$. But this is
just becoming possible with precise space-telescope distance
measurements for supernovae Ia. \\


In pursuing the general integration with these data, we use the framework
and the intermediate results we have presented in Sections \ref{sec:fieqs} and \ref{sec:tddata}. 

First  we note that having determined $\mu_{m_0}(y)$ from data, it is straightforward to obtain 
number counts $N(y)$ from Eq. (\ref{N}). Next, we need the complete determination of the ``potential'' $F(y)$, given by Eq. (\ref{f1}), that at this stage is completely determined from data, except for its dependency on the unknown $\mu_{\Lambda}$. So, our next step is to use Eq. (\ref{MN}) to find the mass function $M(y)$. Then we evaluate the mass function $M(y)$ at $y_{max}$ and plug the result into Eq. (\ref{hellabyeq}) which becomes an algebraic equation for  $\mu_{\Lambda}$. With this determination of $\mu_{\Lambda}$, we know $\dot{C}_0(y)$ completely,
and can now determine $F(y)$ from Eq. (\ref{f1}). Furthermore, we observe from Eq. (\ref{omegadef})  that the quantity $\omega_{0} (y)$  is also completely determined at this stage. 
From here on, we can follow the solution off  $w = w_0$ for all $w$.

It is shown in Ara\'ujo and Stoeger \cite{AS2009b} that from Eqs. (\ref{coneq}) and (\ref{bdd}) 
one obtains the following equation for $\dot{A}(w_0, y)$:

\begin{equation}
\dot{A}_{0}(y) = A_{0}(y)\Biggl\{ \int_0^y\Biggl[2\omega_{0}(\tilde y) +{
\frac 12}\mu_{m_{0}}(\tilde y)\Biggl]{\;A_{0}}^2(\tilde y)\;d{\tilde y} + C_1\Biggr\}. \label{Adplc}
\end{equation}
where, $C_1 = \dot{A}_{0}(0)/A_{0}(0)$. In the {\it Erratum}  of reference \cite{AS2009b}
it is shown that $C_1$ and the subsequent constants $C_n$ that appear in the expressions for higher order derivatives $ \partial^n_w A_0(y)$ (see below) reflect a restricted freedom we
have in choosing the time coordinate $w$ on our world line. We are free to choose these
constants as long as they satisfy the condition of being the coefficients in a Taylor expansion of $A(w,0)$.




Hence, we have shown that the data on our past light cone determines $\dot{A}_{0}(y)$. 
Since we know $\dot{A}_{0}(y)$ from the data,  equations  (\ref{Cdd}) and  (\ref{bdd}) evaluated on our past light cone become algebraic equations for  $\ddot{C}_{0}(y)$  and $\ddot{B}_{0}(y)$, respectively

\begin{eqnarray}
\ddot{C}_{0}(y) &=&  \frac{\dot{C}_0(y){\dot{A}_0}(y)}{A_0(y)}+\Biggl[\omega_{0}(y) +{
\frac 12}\mu_{\Lambda}(y)\Biggl]{\;A_{0}}^2(y)C_0(y) \label{Cddpnc} \\
\ddot{B}_{0}(y) &=& - \frac{A_0^{\prime}(y){\dot{A}_0}(y)}{A_0(y)}-\Biggl[2\omega_{0}(y) +{
\frac 12}\mu_{m_{0}}(y)\Biggl]{\;A_{0}}^2(y)B_0(y)  \label{Bddpnc}
\end{eqnarray}
where in the later we have used Eq. (\ref{coneq}). 

Ara\'ujo and Stoeger \cite{AS2009b} have shown that the next step in this procedure is to 
obtain and equation for $\ddot A_0(w_0,y)$ through differentiation of  Eq. (\ref{bdd}) with 
respect to $w$,  and use of Eq. (\ref{coneq}). That leads to:

\begin{eqnarray}
\ddot A_0(y)&=&-A_0(y) \int^y_0 \Biggl\{\frac{\ddot B_0(\tilde y) \dot A_0(\tilde y)}{A^2_0(\tilde y)} -\frac{\dot B_0(\tilde y)  \dot A^2_0(\tilde y)}{A^3_0(\tilde y)}  \nonumber\\
&  & \hfill{\qquad}  -\frac{1}{A_0(\tilde y)} \biggl\{\frac{\partial}{\partial w}\biggl[ \biggl(2\omega +{\frac 12}\mu_{m} \biggl)\;BA^2\biggl]\biggl\}_0 d \tilde y \Biggr\}-A_0(y)C_2 , \label{Addplc}
\end{eqnarray}
where, $C_2= \ddot A_0(0)/A_0(0)$.

It is important to note that all quantities on the R.H.S. of the above equation are obtainable either directly from the data or from the algorithmic steps in the procedure we are
outlining here [see Ara\'ujo and Stoeger \cite{AS2009b} for a detailed description of the algorithm]. Therefore, we have shown that  we can obtain  $\ddot A_0(y)$ from the data. The repetition of this procedure will give us all time derivatives of $A(w,y)$ $B(w,y)$ and $C(w,y)$ on our past light cone, which means that all the metric functions are completely determined by data on our past light cone, and calculable as Taylor series. 

We see from the above procedure that each step begins by finding the successive time derivatives of the metric function $A(w,y)$ on our past light cone, $\partial^n_wA(w_0,y)$
that is completely determined given our choice of the time coordinate $w$ as measuring proper time along our world line ${\cal C}$.

\section{\label{sec:concl} Conclusion}

We have shown that the mass-energy density of the Universe can be fully determined in 
terms of the cosmological redshift, the redshift-drift  and angular distance (observer area distance). This indirect measurement of the mass-density of the universe is instrumental in overcoming all the well known and highly discussed problems connected to its determination through number counting of sources. That result can also provide a good way of estimating the 
content of cold dark matter in the Universe. 

We have also shown that we can use the redshift-drift as a replacement for the mass-energy density element in the minimally required data set to construct an LTB model for the Universe in observational coordinates. In fact, this use of the cosmological redshift drift as a fundamental piece of data simplifies the integration procedure of the field equations in observational coordinates through the use of the much simpler Eq. (\ref{jacob0}) to find  $z(y)$ instead of the more complex null Raychaudhuri Eq. (\ref{nr}) that in this new scheme, becomes a consistency relation. 








\end{document}